
\magnification=\magstep1
\input amstex
\documentstyle{amsppt}
\overfullrule=0pt
\TagsOnLeft
\CenteredTagsOnSplits

\def\R{{\Bbb R}}
\def\C{{\Bbb C}}
\def\Z{{\Bbb Z}}
\def\P{{\Bbb P}}

\def\O{{\Cal O}}

\def\h{{\frak h}}
\def\g{{\frak g}}
\def\gh{{\widehat{\frak g}}}

\def\a{{\frak a}}

\def\M{{\Cal M}}
\def\J{{\Cal A}}
\def\pp{w}
\def\t{x}
\def\tP{\wp}

\def\E{{\Cal E}}

\def\l{\ell}

\def\H{{\Cal H}}
\def\Hl{{\Cal H_{\lm}}}

\def\Hv{\H_{\vec \lm}}
\def\F{{\Cal F}}

\def\f{{\frak F}}
\def\vo{\varphi}
\def\Vo{\varphi}

\def\vlm{{\vec \lm}}
\def\da{^\dagger}
\def\Vac{{\Cal V}}
\def\E{{\Cal E}}
\def\cals{{\Cal S}}

\def\plm{\frac{\lm}{2},\frac{\lm+1}{2},\ldots,\frac{\lm+L}{2}}

\def\X{{\frak X}}
\def\Th{\Theta}

\def\-{.}

\def\lm{\lambda}

\def\zN{z_1,\ldots,z_N}
\def\vz{\hbox{\bf z}}
\def\al{\alpha}
\def\Cx{\C ((\xi))}

\def\<{\langle}
\def\>{\rangle}
\def\PH{\<\Phi}

\def\Tr{\hbox{\rm Tr}_{\H_\mu} \vo}

\def\tr{\hbox{\rm Tr}_{\H_\mu}}
\def\uk{|u^k\>}

\def\q{q^{L_0-\frac{c_v}{24}}}
\def\pp{\xi}
\def\p{\pp^{\frac{H}{2}}}

\def\pap{\pp\pa_{\pp}}
\def\paq{q\pa_{q}}
\def\paz{z\pa_{z}}
\def\pazj{\pa_{z_j}}
\def\pat{t\pa_{t}}

\def\DT{\hbox{\bf Det}}
\def\Hom{\hbox{\rm Hom}\,}
\def\Res{\hbox{\rm Res}\,}
\def\Hom{\mathop{\hbox{\rm Hom}}\nolimits}

\def\pa{\partial}
\rightheadtext{the SU(2) WZNW model on elliptic curves}
\topmatter
\title
  differential equations associated to \\
the SU(2) wznw model on elliptic curves
\endtitle
\author  Takeshi Suzuki  \endauthor
\affil
        Research Institute for Mathematical Sciences,
        Kyoto University,
        Kyoto, 606-01,
        Japan\\
        e-mail : takeshi\@kurims.kyoto-u.ac.jp
\endaffil

\abstract
We study the $SU(2)$ WZNW model over a
family of elliptic curves.
Starting from the formulation
 developed in [TUY],
we derive a system of differential equations
which contains the Knizhnik-Zamolodchikov-Bernard equations[Be1][FW].
Our system completely determines the $N$-point functions and
is
regarded as a natural elliptic analogue of
the system obtained in [TK] for the
projective line.
We also calculate the system for
the 1-point functions explicitly.
This gives a generalization of the results in
[EO2] for $\widehat{\frak sl}(2,\C)$-characters.
\endabstract
\endtopmatter
\document
\subhead
{\S0. Introduction}
 \endsubhead
\medskip
We consider the Wess-Zumino-Novikov-Witten (WZNW) model.
A mathematical formulation of this model on general algebraic curves
 is given in [TUY],
where the correlation functions
are defined as flat sections of a certain vector bundle
 over the moduli space of curves.
On the projective line $\P^1$,
the correlation functions are realized more explicitly
in [TK] as functions which take
their values in a certain finite-dimensional vector space,
and characterized by  the system of equations containing
the Knizhnik-Zamolodchikov (KZ) equations[KZ].
One aim in the present paper is to have a parallel description
on elliptic curves.
Namely, we characterize the $N$-point functions as vector-valued functions
by a system of differential equations containing
an elliptic analogue of the KZ equations by Bernard[Be1].
Furthermore we
write down this system explicitly in the 1-pointed case.

To explain more precisely,
first let us review the formulation in [TUY] roughly.
Let $\g$ be a simple Lie algebra over $\C$
and $\gh$ the corresponding affine Lie algebra.
We fix a positive integer $\l$ (called the level) and consider the
integrable highest weight modules of $\gh$ of level $\l$.
Such modules are parameterized by the set of highest weight
$P_\l$ and we denote by $\H_\lm$
 the left module corresponding to
$\lm\in P_\l$.
By $M_{g,N}$ we denote the moduli space of $N$-pointed curves
of genus $g$.
For $\X\in M_{g,N}$ and
$\vlm=(\lm_1,\ldots,\lm_N)\in (P_\l)^N$, we associate
the space of conformal blocks $\Vac_g\da(\X;\vlm)$.
The space $\Vac_g\da(\X;\vlm)$ is the finite dimensional subspace of
$\H_\vlm\da:=\Hom_\C(\H_{\lm_1}\otimes\cdots\otimes\H_{\lm_N},\C)$
 defined by ``the gauge conditions''.
Consider the vector bundle
$\tilde{\Vac}_g\da(\vlm)=\cup_{\X\in M_{g,N}}\Vac_g\da(\X;\vlm)$
over ${M}_{g,N}$.
On this vector bundle, projectively flat connections are defined through
the Kodaira-Spencer theory, and flat sections of $\tilde{\Vac}_g\da(\vlm)$
with respect to these connections are called the $N$-point correlation
functions
(or $N$-point functions).
In the rest of this paper we set $\g=\frak{sl}(2,\C)=\C E\oplus\C F\oplus \C H$
 for simplicity, where $E,F$ and $H$ are the basis of $\g$ satisfying
$$[H,E]=2E,\ [H,F]=-2F,\ [E,F]=H.$$
We identify $P_\l$ with the set $\{\; 0,\frac{1}{2},\ldots
,\frac{\l}{2}\;\}$ by the mapping $\lm\mapsto \frac{\lm(H)}{2}$.

In the case of genus 0,
the space of conformal blocks is injectively mapped into
$V_{\vlm}\da:=
\Hom_\C(V_{\lm_1}\otimes\cdots\otimes V_{\lm_N},\C)$
by the restriction map, where $V_\lm\subset \H_\lm$ denotes the
finite dimensional
irreducible highest weight
left $\g$-module with highest weight $\lm$.
This injectivity
makes it possible to treat this model in a more explicit way as above,
and the $N$-point functions are described by the vacuum expectation values
of vertex operators.

On the other hand, in the case of genus 1 this injectivity does not hold,
and in order to recover it
we twist the space of conformal blocks
by introducing a new parameter following [Be1,2][EO1][FW].
Because of the twisting, any $N$-point function in genus 1
can be calculated from its restriction to $V_\vlm$ (Proposition 3.3.2).
It is natural to ask how the restrictions of the $N$-point functions
 are characterized as  $V_\vlm\da$-valued functions.
It turns out that the restricted $N$-point functions
satisfy the equations (E1)--(E3) in Proposition 3.3.3.
These equations are essentially derived by Bernard[Be1]
for traces of vertex operators
$$\hbox{Tr}_{\H_\mu}(
 \varphi_1(z_1)\cdots\varphi_N(z_N)q^{L_0-\frac{c_v}{24}}\p)
\in V_\vlm\da,$$
where
$z_1,\ldots,z_N,q,\pp$
are the variables in $\C^*$ with $|q|<1$,
$\varphi_j: V_{\lm_j}\otimes \H_{\mu_{j}}\to\hat\H_{\mu_{j-1}}$
 $(j=1,\ldots,N)$
 are the vertex operators for some $\mu_i\ (i=0,\ldots,N)$ with
$\mu_0=\mu_N=\mu$, $L_0$
 is defined in (1.2.1) and $c_v=3\l/(\l-2)$
(for the details, see \S\S 3.4).
It is proved that the space of restricted $N$-point functions
is spanned by traces of vertex operators (Theorem 3.4.3)
and hence Bernard's approach is equivalent to ours.
However, the system (E1)--(E3) is not complete since
it has infinite-dimensional solution space.

We will show that the integrability condition
$$\left(E\otimes t^{-1}\right)^{\l-2\lm+1}|\bar{v}(\lm)\>=0$$
for the highest weight vector  $|\bar{v}(\lm)\>\in\H_\lm$
implies the differential equations (E4), which determine
the $N$-point functions completely combining with (E1)--(E3).

For 1-point functions, the equation (E4) can be written down explicitly,
and the system (E1)--(E4) reduces to the two equations (F1)(F2)
in Theorem 4.2.4.
In the simplest case, the 1-point functions are given by the characters
$$\hbox{\rm Tr}_{\H_\mu}\q\p\ \ \ \left(\mu=0,\frac{1}{2},\ldots,\frac{\l}{2}
\right),$$
and our system coincide with the one obtained in [EO2].

Recently Felder and Wieczerkowski
give a conjecture on
the characterization of the restricted $N$-point functions in genus 1
by using the modular properties and certain additive conditions[FW].
They confirm their conjecture in some cases by explicit calculations.
We partly recover this result
by solving the equation (F2) (Proposition 4.2.5).
The equation (F1) can be also integrated when
 the dimension of the solution space is small, and we can
calculate the 1-point functions explicitly.

\subhead
\S 1. Representation theory for $\widehat{\frak{sl}}(2,\C)$
\endsubhead
\medskip
For the details of the contents in this section, we refer the reader to [Kac].
\medskip
{\bf 1.1 Integrable highest weight modules.}
\medskip
By $\C [[\t]]$ and $\C ((\t))$, we mean the ring of formal power series
in $\t$ and the field of formal Laurent series in $\t$, respectively.
\noindent
We put $\g={\frak {sl}}(2,\C) $.
Let
$\h=\C H$ be a Cartan subalgebra of $\g$ and
$
(\ ,\ ) : \g \times \g \to \C
$
 the Cartan-Killing form normalized by the condition
$(H,H)=2$.
We identify the set $P_+$ of dominant integral weights with
$\frac{1}{2}\Z_{\geq0}$.
For $\lm\in P_+$,
we denote by $V_\lm$ the irreducible highest weight
left $\g$-module with highest weight $\lm$ and by $|v(\lm)\>$
its highest weight vector.

The affine Lie algebra $\gh$ associated with $\g$ is defined by
$$
\gh = \g\otimes\C((\t))\oplus\C c,
$$
where $c$ is an central element of $\gh$
and the Lie algebra structure is given by
$$
[X\otimes f(\t),Y\otimes g(\t)] = [X,Y]\otimes f(\t)g(\t) +
c\cdot (X,Y)\, \mathop{\Res}_{\t = 0} (g(\t)\cdot df(\t)),
$$
for
$X,Y \in \g,\  f(\t),g(\t) \in \Cx.$
We use the following notations:
$$\gather
X_n = X \otimes \t^n,\ X = X_0\, ,\\
\gh_+ =\g\otimes\C [[\t]]\t,\ \gh_- =\g\otimes\C [\t^{-1}]\t^{-1},\\
\widehat{\frak p}_\pm =\gh_\pm\oplus\g\oplus\C c.
\endgather
$$

Fix a positive integer $\ell$ (called the level) and put
$
P_\l = \{\; 0,\frac{1}{2},\ldots,\frac{\ell}{2}\; \}\subset P_+.
$
For $\lm\in P_\l$, we define the action of
$\widehat{\frak p}_+$ on $V_\lm$ by
$c=\l\times id$ and $a=0$ for all $\a\in\gh_+$,
and put
$$\M_\lm=U(\gh)\otimes_{\widehat{\frak p}_+} V_\lm.$$
Then $\M_\lm$ is a highest weight
left $\gh$-module and it has the maximal proper submodule ${\Cal J}_\lm$,
which is generated by the singular vector $E_{-1}^{\l-2\lm+1}|v(\lm)\>$:
$${\Cal J}_\lm=U(\widehat{\frak p}_-)E_{-1}^{\l-2\lm+1}|v(\lm)\>.$$
The integrable highest weight left $\gh$-module $\H_\lm$ with
 highest weight $\lm$ is
defined as the quotient module $\M_\lm / {\Cal J}_{\lm}$.
We denote by $|\bar{v}(\lm)\>$  the  highest weight
vector in $\H_\lm$.
We introduce the lowest weight right $\gh$-module structure on
$$\H_{\lm}^\dagger=\Hom_\C(\H_\lm,\C)$$
in the usual way,
and denote its lowest weight vector by $\<\bar{v}(\lm)|$.
\medskip
{\bf 1.2. Segal-Sugawara construction and the filtration on $\H_\lm$.}
\medskip
Fix a weight $\lm\in P_\l$.
On $\Hl$, elements $L_n\ (n\in\Z)$ of the Virasoro algebra
act through the Segal-Sugawara construction
$$
L_n\ =\ \frac{1}{2(\ell+2)}\sum_{m\in\Z}\left\{
{}_{\circ}^{\circ}\frac{1}{2}H_m H_{n-m}{}_{\circ}^{\circ}
+{}_{\circ}^{\circ}E_m F_{n-m}{}_{\circ}^{\circ}
+{}_{\circ}^{\circ}F_m E_{n-m}{}_{\circ}^{\circ}\right\} ,\tag 1.2.1
$$
where
${}_\circ^\circ\ {}_\circ^\circ$ denotes the standard normal ordering,
and the operators $L_n\ ( n\in\Z)$ satisfy the following commutation relations:
$$
\align
[L_n,L_m]\ & =\ (n-m)L_{n+m}\ +\ \frac{c_v}{12} (n^3-n)\delta_{n+m,0}\, , \\
[L_n,X_m]\ & =\ -mX_{n+m}\hbox{ for }X\ \in\ \g\, ,
\endalign
$$
where $c_v$ is the central charge of Virasoro algebra:
$$
c_v\ =\ \frac{3\l}{\l+2}.\tag 1.2.2
$$

Put
$$
X(z) = \sum_{n\in\Z}X_nz^{-n-1}\ \ (X\in\g),\
T(z) = \sum_{n\in\Z}L_nz^{-n-2}.
$$

The module
$\H_\lm$ has the decomposition
$\H_\lm=\oplus_{d\geq 0}\H_\lm(d)$,
where
$$\gather
\H_\lm(d)=\{\; |u\>\in\H_\lm\; ;\; L_0|u\>=(\varDelta_\lm+d)|u\>\; \},\\
\varDelta_\lm=\frac{\lm(\lm+1)}{\l+2}.\tag 1.2.3
\endgather
$$

We define the
filtration $\{\F_\bullet\}$ on $\H_\lm$ by
$$\F_p\H_\lm=\sum_{d\le p}\H_\lm(d)$$
 and put
$
\hat{\H}_\lm =\prod_{d\geq 0}\H_\lm (d)\, .\;
$
\medskip
{\bf 1.3. The Lie algebra $\gh_N$.}
\medskip
Put $L\g=\g\otimes\C((\t))$.
For a positive  integer $N$, we define a Lie algebra $\gh_N$
by
$$
 \gh_N= \oplus_{j=1}^N L\g_{(i)}\oplus \C c,
$$
where $L\g_{(i)}$ denotes a copy of $L\g$ and $c$ is a center.
The commutation relations are given by
$$\align
[& \oplus_{j=1}^N X_j\otimes f_j,\oplus_{j=1}^N Y_j\otimes g_j]\ = \\
& \oplus_{j=1}^N [X_j,Y_j]\otimes f_jg_j\ +\ \sum_{j=1}^N (X_j,Y_j)
\mathop{\Res}_{\xi_j=0} (g_j\cdot df_j)\cdot c\, .
\endalign
$$
For each $\vec\lm= (\lm_1,\ldots,\lm_N)\in (P_\l)^N$ a left
$\gh_N$-module $\Hv$ is defined by
$$
\Hv = \H_{\lm_1}\otimes\cdots\otimes\H_{\lm_N}\, .
$$
Similarly a right $\gh_N$-module $\H_\vlm^\dagger$ is defied by
$$
\Hv^\dagger  = \H_{\lm_1}^\dagger
\hat{\otimes}\cdots\hat{\otimes}\H_{\lm_N}^\dagger
\cong\Hom_\C(\Hv,\C).
$$
The $\gh_N$-action on $\Hv$ is given by
$$
\align
c&=\l\cdot id\\
(\oplus_{j=1}^N a_j)|u_1\otimes\cdots\otimes u_N\>&=
\sum_{j=1}^N \rho_j(a_j)|u_1\otimes\cdots\otimes u_N\>
\endalign
$$
for $a_j\in L\g_{(j)} (j=1,\ldots,N)$, where we used the
notations
$$\align
&|u_1\otimes\cdots\otimes u_N\>=|u_1\>\otimes\cdots\otimes|u_N\>,\\
&\rho_j(a)|u_1\otimes\cdots\otimes u_N\>=|u_1\otimes\cdots
\otimes a\cdot u_j\otimes\cdots\otimes u_N\>
\endalign
$$
for $|u_i\>\in\H_{\lm_i}\, (i=1,\ldots,N)$ and $ a\in L\g $.
The right action on $\Hv^\dagger$ is defined similarly.
The module $\H_\vlm$  has the filtration
 induced from those of
$\H_{\lm_j}\ (j=1,\ldots,N)$:
$$
\F_p\H_\vlm=\sum_{d\le p}\H_\vlm(d)\, ,
$$
where
$$\H_\vlm(d)=\sum_{d_1+\cdots+d_N=d}\H_{\lm_1}(d_1)\otimes
\cdots\otimes\H_{\lm_N}(d_N).
$$
We put
$$\align
V_\vlm&=V_{\lm_1}\otimes\cdots\otimes V_{\lm_N}
\cong \Hv(0) ,\\
V_\lm\da&=\Hom_\C(V_\vlm,\C).
\endalign$$

 \subhead
 {\S 2 The WZNW model in genus 0}
 \endsubhead
\medskip
In this section we review the SU(2) WZNW model on the projective line $\P^1$.
\medskip
{\bf 2.1. The space of conformal blocks.}
\medskip
In this subsection
we define the $N$-point functions
on $\P^1$ following [TUY] as
sections of a vector bundle on the manifold
$$
R_N=\{\; (\zN)\in (\C^*)^N\; ;\; z_i\neq z_j \hbox{ if } i\neq j\; \}.
$$
For a meromorphic function $f(t)$ on $\P^1$ and $w\in\C$,
put
$$
\align
X[f(t)]_w     & = \mathop{\Res}_{t=w} f(t)X(t-w)dt\, ,\\
T[f(t)\frac{d}{dt}]_w &= \mathop{\Res}_{t=w} f(t)T(t-w)dt\, .
\endalign
$$
If $f(t)$ has an Laurent expansion $f(t)=\sum_{n\geq M}a_n(t-w)^n$
then  $X[f(t)]_w$ is an element of $\gh$ given by
$$X[f(t)]_w=\sum_{n\geq M}a_n X_n.$$
For $z=(\zN)\in R_N$, we set
 $$\gh(z)=H^0(\P^1,\g\otimes\O_{\P^1} (*\sum_{j=1}^{N}z_j)).$$
Then
we have the following injection:
$$
\align
 \gh(z)\ \ &\to\  \ \gh_N\, , \\
X\otimes f(z)\ &\mapsto\ X[f]:=\oplus_{j=1}^N X[f]_{z_j}.
\endalign
$$
Through this map we regard $\gh(z)$ as a subspace of $\gh_N$ and
the residue theorem implies
that $\gh(z)$ is a Lie subalgebra of $\gh_N$.
We also use the following notation
$$
T[g]=\oplus_{j=1}^N T[g]_{z_j}
$$
for $g\in H^0(\P^1,\Th_{\P^1} (*\sum_{j=1}^{N}z_j))$, where
$\Th_{\P^1}$ denotes the sheaf of vector fields on $\P^1$.
\medskip
\noindent
{\bf Definition 2.1.1.}
For $z=(\zN)\in R_N$ and $\vlm=(\lm_1,\ldots,\lm_N)\in (P_\l)^N$
we put
$$
\align
\Vac_0(z\, ;\vlm)\ & =\ \Hv\ /\gh (z) \Hv ,\\
\Vac_0^\dagger (z\, ;\vlm)\ & =\ \{\ \<\Psi|\in\Hv^\dagger\ ;\
\<\Psi|\gh(z)=0\ \}\\
&\cong \Hom_\C(\Vac_0(z\, ;\vlm),\C)\, .
\endalign
$$
We call $\Vac_0^\dagger(z\, ;\vlm)$ the space of conformal blocks
(or the space of vacua)
in genus 0 attached to $(z\, ;\vlm)$.
\medskip
For a vector space $V$ and a complex manifold $M$, we denote
by $V[M]$ the set of multi-valued, holomorphic $V$-valued functions on $M$.
\medskip
\noindent
{\bf Definition 2.1.2.}
For $\vlm\in (P_\l)^N$,
an element
$\PH|$ of $\H_\vlm\da[R_N]$ is called an $N$-point function in genus 0
attached to $\vlm$
if the following conditions are satisfied:
\medskip
(A1) For each $z\in R_N$,
$$
\PH(z)|\in\Vac_0\da(z;\vlm)
$$
\medskip
(A2) For $j=1,\ldots,N$,
$$\pa_{z_j} \PH(z)|=\PH(z)|\rho_j(L_{-1}).$$
By $\f_0(\vlm)$ we denote the set of $N$-point functions in genus 0
attached to $\vlm$.
\medskip

\remark
{Remark}
 The condition (A1) implies the following:
\medskip
(A1$'$) For each $z\in R_N$,
$$\PH(z)|T[g]=0$$
for any $ g\in H^0(\P^1,\Th_{\P^1}(*\sum_{j=1}^N z_j))$.
\endremark
\medskip
\medskip
{\bf 2.2. Restrictions of the $N$-point functions to $V_\vlm$.}
\medskip
A remarkable property of the space of conformal blocks in  genus 0 is
the following:
\proclaim
{Lemma 2.2.1}
The  composition map
$$V_\vlm\ \hookrightarrow\ \H_\vlm\ \to\ \Vac_0(z\, ;\vlm)$$
is surjective. In other words, the restriction map
$$\Vac_0^\dagger(z\, ;\vlm)\ \to\ V_\vlm^\dagger$$
is injective.
\endproclaim
This lemma implies that,
for an $N$-point function $\PH|$, we can calculate
$\<\Phi|u\>$ for any $|u\>\in\H_\vlm$, from the data
$\{\;\<\Phi|v\>$ ; $|v\>\in V_\vlm\;\}$.
By $\f_0^r(\vlm)$ we denote the image of $\f_0(\vlm)$ in $V_\vlm\da[R_N]$
under the restriction map.
It is natural to ask how the set
$\f_0^r(\vlm)$ is characterized
in $V_\vlm\da[R_N]$,
and the answer is given as follows:
\proclaim
{Proposition 2.2.2}{\rm [TK]}
The space $\f_0^r(\vlm)$ coincides with
 the solution space of the following system
of  equations:
\medskip

{\rm (B1)} For each $X\in\g$,
$$\sum_{j=1}^N\<\phi(z)|\rho_j(X)=0.$$

{\rm (B2) [the Knizhnik-Zamolodchikov equations]}
For each $j=1,\ldots,N$,
$$ (\l+2)\pa_{z_j} \<\phi(z)|=\sum_{i\neq j}\<\phi(z)|
\frac{\Omega_{i,j}}{z_i-z_j}\, ,$$
where
$$\Omega_{i,j}=
\frac{1}{2}\rho_i(H) \rho_j(H)+ \rho_i(E)  \rho_j(F)+
\rho_i(F) \rho_j(E)\, .
$$

{\rm (B3)} For each $j=1,\ldots,N$,
$$\align \sum_{n_1+\cdots +n_N=\l_j}
&\binom{\l_j}{\vec{n}_j}
\prod_{i\neq j}(z_i-z_j)^{-n_i}\<\phi(z)|E^{n_1}v_1\otimes\cdots\otimes
v(\lm_j)\otimes\cdots\otimes E^{n_N}v_N\>\\
&=0
\endalign
$$
for any $|v_i\>\in V_{\lm_i}\ (i\neq j)$.
Here $\l_j=\l-2\lm_j+1$, $\vec{n}_j=(n_1,\ldots,n_{j-1},n_{j+1},\ldots,n_N)$
and
  $\binom{\l_j}{\vec{n}_j}$ is the multinomial coefficient. $\square$
\endproclaim
\remark
{Remark}
The equation (B3) is a consequence of the integrability condition
$$E_{-1}^{\l-2\lm_j+1}|\bar{v}(\lm_j)\>=0 \ (j=1,\ldots,N),\tag 2.2.1$$
for the highest weight vector $|\bar{v}(\lm_j)\>\in\H_{\lm_j}$.
\endremark
\medskip
{\bf 2.3. Vertex operators.}
\medskip
We review the description of $N$-point functions by vertex operators.
\medskip\noindent
{\bf Definition 2.3.1.}
For $(\nu,\lm,\mu)\in (P_\l)^3$
a multi-valued, holomorphic, operator valued function $\vo (z_1)$
on the manifold $\C^*=\C\setminus \{0\}$ is called a vertex operator of
type $(\nu,\lm,\mu)$, if
$$
\vo (z_1)\ :\ V_{\lm}\otimes \H_{\mu} \to \hat{\H}_{\nu}
$$
satisfies the following conditions:
\medskip
(C1) For $X\in\g,\; |v\>\in V_\lm$ and $ m\in\Z ,$
$$[X_m,\vo (|v\>;z_1)]\ =\ z_1^m\vo(X|v\>;z_1).$$
\medskip
(C2) For $|v\>\in V_\lm$ and $m\in\Z, $
$$[L_m, \vo (|v\>;z_1)]\ =\ z_1^m\left\{z_1\frac{d}{dz_1}+(m+1)
\varDelta_\lm\right\}\vo (|v\>;z_1).$$
\medskip
Here
$\vo(|u\>;z_1):\H_{\nu} \to \hat{\H}_{\mu} $
is the operator defined by
$\vo(|u\>;z_1)|v\>=\vo(z_1)|u\otimes v\>$
for $|u\>\in V_\lm$ and $|v\>\in \H_{\nu}$.
\medskip

For vertex operators $\vo_j (z_j)$
$(j=1,\ldots,N)$,
the composition $\vo_1 (z_1)\cdots\vo_N (z_N)$ makes sense for
$|z_1|>\cdots >|z_N|$
 and analytically continued to $R_N$.
\proclaim
{Proposition 2.3.2}{\rm [TK]}
The space $\f_0^r(\vlm)$ is spanned by the
following $V_\vlm\da$-valued functions:
$$\<v(0)|\vo_1 (z_1)
\cdots\vo_N (z_N)|v(0)\>\, ,
$$
where $\vo_j\, (j=1,\ldots,N)$ is the vertex operator of type
$(\mu_{j-1},\lm_j,\mu_{j})$ for some $\mu_i\in P_\l$ $(i=0,\ldots,N)$ with
$\mu_0=\mu_N=0$.
\endproclaim
\medskip
\proclaim
{Proposition 2.3.3}\hbox{\rm [TK]}
Any nonzero  vertex operator
$$\varphi(z_1):V_\lm\otimes\H_{\mu}\to\hat\H_{\nu}$$
is uniquely extended to the operator
$$\hat\varphi(z_1):\M_\lm\otimes\H_{\mu}\to\hat\H_{\nu}$$
by the following condition:
$$
\hat\vo(X_n|u\>;z_1)=\mathop{\Res}_{w=z}(w-z_1)^n\hat\vo(|u\>;z_1)X(w)dw,\tag
2.3.1
$$
for each $|u\>\in\M_\lm$, $X\in\g$ and $n\in\Z$.
\medskip
Moreover, $\hat{\varphi}$ has the following properties:
$$\gather
\pa_z\hat\varphi(|u\>;z_1)=\hat\varphi(L_{-1}|u\>;z_1)\ \ \ \hbox{ for any }
|u\>\in\M_\lm,\tag 2.3.2\\
\hat\varphi (|u\>;z_1)=0\ \ \ \hbox{ for any }
|u\>\in {\Cal J}_\lm= U(\widehat{\frak p}_-)E_{-1}^{\l-2\lm+1}|{v}(\lm)\>.\tag
2.3.3
\endgather
$$
\endproclaim

The property (2.3.3) implies that $\hat\varphi$ reduces to the operator
$$\hat\varphi(z_1):\H_\lm\otimes\H_{\mu}\to\hat\H_{\nu}.$$

\subhead
{\S 3 The WZNW model in genus 1}
\endsubhead
\medskip
In this section we consider the elliptic analogue of the story
in the previous section.
Our aim is to embed the set of $N$-point functions in genus 1
(Definition 3.1.3) into the set of $V_\vlm\da$-valued functions,
and to characterize its image by a system of differential equations.
We also show that the $N$-point functions are given by the traces of vertex
 operators.
\medskip
{\bf 3.1 Functions with quasi-periodicity.}
\medskip
First, we prepare some functions for the later use.
Put $D^*=\{\; q\in \C^*\; ;\; |q|<1\;\}$ and introduce the following functions
on $\C^*\times D^*$:
$$
\align
&\Th(z,q)  =\sum_{ n\in\Z+\frac{1}{2} } (-1)^{n+1}q^{\frac{1}{2} n^2}z^n
\tag 3.1.1\\
          &=-\sqrt{-1}z^{\frac{1}{2}}q^{\frac{1}{8}}\prod_{n\geq 1}(1-q^n)
          (1-z q^n )(1-z^{-1}q^{n-1})\, ,    \\
&\zeta(z,q)=\frac{\paz \Th(z,q)}{\Th(z,q)}.                \tag 3.1.2
\\
&\wp(z,q)  =-\paz\zeta(z,q)+2\frac{\paq\eta(q)}{\eta(q)},
                                \tag 3.1.3
\endalign
$$
where $\eta(q)$ is the Dedekind eta function
$$\eta(q)=q^{\frac{1}{24}}\prod_{n\geq1}(1-q^n).$$
The function $\Th(z,q)$ satisfies the heat equation
$$2\paq\Th(z,q)=(\paz)^2\Th(z,q).$$
The function $\wp(z,q)$ satisfies
$\wp(qz,q)=\wp(z,q),$
and $\zeta(z,q)$ have the following quasi-periodicity:
$$\zeta(qz,q)=\zeta(z,q)-1.\tag 3.1.4$$
For $(z,q)\in \C^*\times D^*$ and $\pp\in\C^*$, we put
$$
\sigma_{\pm}(z,q,\pp) =
\frac{\Th(z^{-1}\pp^{\pm 1},q)\Th'(1,q)}{\Th(z,q)\Th(\pp^{\pm 1},q)}
\tag 3.1.5
$$
Here $\Th'(z,q)=\paz\Th(z,q)$.
The function $\sigma_{\pm}(z,q,\pp)$ have
 the following properties:
$$\split
\sigma_{\pm}(qz,q,\pp)&=\pp^{\pm 1}\sigma_{\pm}(z,q,\pp)\, ,\\
\sigma_{\pm}(z^{-1},q,\pp) &= -\sigma_{\mp}(z,q,\pp)\, .
\endsplit \tag3.1.6
$$
For $\zeta(z,q)$ and $\sigma_\pm(z,q,\pp)$, we have the following expansion
at $z=1$:
$$
\align
&\zeta(z,q)=\frac{1}{z-1}+\frac{1}{2}-2\al(q)(z-1)+O(z-1)^2,\tag 3.1.7\\
&\sigma_{\pm}(z,q,\pp)=\frac{1}{z-1}\mp\zeta(\pp,q)+\frac{1}{2}\tag 3.1.8\\
&-\sum_{n\geq1}\left(
 \frac{n\pp^{-1}q^n}{1-\pp^{-1}q^n}
+\frac{n\pp q^n}    {1-\pp q^n}
\right)(z-1)+O(z-1)^2,
\endalign
$$
where $\al(q)$ is given by
$$\al(q)=-\frac{\paq\eta(q)}{\eta(q)}+\frac{1}{24}.
\tag 3.1.9$$
\medskip
{\bf 3.2. Twisting the space of conformal blocks.}
\medskip
In the case of genus $1$ (or $>0$), if we work with
 the formulation of [TUY],
an $N$-point function is not determined by its restriction
on $V_\vlm$.
In order to resolve this difficulty we ``twist'' the space of
conformal blocks following [Be1,2][EO1][FW].

For $q\in D^*$, we consider the elliptic curve
$
\E_q\ =\ \C^*/\<q\>,
$
where $\<q\>$ is the infinite cyclic group of automorphisms generated by
$
z \mapsto qz.
$
We denote by $[z]_q$ the image of a point $z\in\C^*$ on $\E_q$
and  put
$$T_N=\{\; (z,q)=(\zN,q)\in (\C^*)^N\times D^*\; ;\; [z_i]_q\neq [z_j]_q
\hbox{ if }i\neq j\;\}.$$
In the following we omit the subscript $q$ in $[z]_q$.
For $(z,q)\in T_N$ and $\vlm=(\lm_1,\ldots,\lm_N)\in P_\l$
we can define the space of conformal blocks
attached to the elliptic curve $\E_q$:
$$\Vac_1^\dagger([z],q ;\vlm)=\{\ \<\Psi|\in\H_\vlm\da\;
;\; \<\Psi|\g([z],q)=0\ \},$$
where
$$\gh([z],q)= H^0(\E_q,\g\otimes\O_{\E_q}(*\sum_{j=1}^N [z_j])),$$
but for our purpose we need to twist it as follows.
We introduce a new variable $\pp\in \C^*$, and put
$$
\gh([z],q,\pp) = \left\{\; a(t)\in
H^0(\C^*,\g\otimes\O_{\C^*}(*\sum_{j=1}^{N}\sum_{n\in\Z}q^n z_j))\ ;
\ a(qt) = \p(a(t))\pp^{-\frac{H}{2}}\; \right\}.
$$
This space is regarded as the space of meromorphic sections of
the $\g$-bundle which is twisted by $\p$ along the cycle
$\{\ [w]\in \E_q\; ;\; w\in\R, q\le w<1\ \}$.
For $\pp=1$, we have
$$\gh([z],q,1)=H^0(\E_q,\g\otimes\O_{\E_q}(*\sum_{j=1}^{N}[z_j])).$$
%
As in the previous section
we have the following injection:
$$\align
 \gh([z],q,\pp)&\to \ \gh_N \\
 X\otimes f\; &\mapsto \ X[f].
\endalign
$$
By this map we regard $\gh([z],q,\pp)$
 as a subspace of $\gh_N$.
Furthermore we can easily have the following lemma.
\proclaim
{Lemma 3.2.1}
The vector space $\gh([z],q,\pp)$ is a Lie subalgebra of $\;\gh_N$. $\square$
\endproclaim
\medskip
\noindent
{\bf Definition 3.2.2.}
Put
$$
\align
\Vac_1   ([z],q,\pp;\vlm)&=\H_\vlm / \gh([z],q,\pp)\H_\vlm\, ,\\
\Vac_1\da([z],q,\pp;\vlm)&=\{\ \<\Psi|\in\H_\vlm^\dagger\ ;\
\<\Psi|\gh([z],q,\pp)=0\ \}\\
&\cong\ \Hom_\C(\Vac_1([z],q,\pp;\vlm),\C)\, .
\endalign
$$
We call $\Vac_1^\dagger([z],q,\pp;\vlm)$ the space of conformal blocks
in genus 1 attached to $([z],q,\pp;\vlm)$.
\medskip
Following [TUY][FW], we define the N-point functions in genus 1 as follows:
\medskip\noindent
\definition
{Definition 3.2.3}
An element $\PH|$ of $\H_\vlm^\dagger[T_N\times \C^*]$
 is called an $N$-point function
in genus 1 attached to $\vlm$ if the following conditions are satisfied:

\medskip
(D1) For each $(z,q,\pp)\in T_N\times\C^*$,
$$\PH(z,q,\pp)|\in\Vac_1^\dagger ([z],q,\pp;\vlm).$$

(D2) For $j=1,\ldots,N$,
$$\pazj \PH(z,q,\pp)|=\PH(z,q,\pp)|\rho_j(L_{-1})
$$

(D3)
$$\left(\paq+\frac{c_v}{24}\right) \PH(z,q,\pp)|
=\PH(z,q,\pp)|T\left[\zeta(t/z_1,q)t\frac{d}{dt}\right],$$
 where $\zeta(t,q)$ is the function given by (3.1.2).
\medskip
(D4)
$$\pap \PH(z,q,\pp)|=\PH(z,q,\pp)|\frac{1}{2}H[\zeta(t/z_1,q)].$$
\medskip
We denote by $\f_1(\vlm)$ the set of $N$-point functions attached to $\vlm$.
\enddefinition
\medskip

\remark
{Remark}
(i) The condition (D1) implies the following:
\medskip
(D1$'$) For each $(z,q,\pp)\in T_N\times\C^*$,
$$\<\Phi(z,q,\pp)|T[g]=0$$
for any $g\in H^0(\E_q, \Th_{\E_q}(*\sum_{j=1}^N z_j))
= H^0(\E_q, \O_{\E_q}(*\sum_{j=1}^N z_j)t\frac{d}{dt}).$
\medskip

(ii) The equations (D1)--(D4)
 are compatible with each other due to (3.1.4), e.g.
$$\align
&\left[\pap-\frac{1}{2}H[\zeta(t/z_j)]\,  ,\, X[f(t,q,\pp)]\right]=\tag 3.2.1\\
X[\pap &f(t,q,\pp)]-\frac{1}{2}[H,X]
[\zeta(t/z_j,q)f(t,q,\pp)]\in\gh([z],q,\pp).
\endalign
$$
for $X[f]\in\gh([z],q,\pp)$.
Conversely, the compatibility condition demands (3.1.4) for $\zeta$.

(iii)
In (D3) and (D4) we can replace $\zeta(t/z_1,q)$ with
$\zeta(t/z_j,q)\ (j=2,\ldots,N)$
provided (D1) since
$$\zeta(t/z_1,q)-\zeta(t/z_j,q)\in H^0(\E_q,\O_{\E_q}(*\sum_{i=1}^{N}z_j)).
\tag 3.2.2$$
\endremark
\medskip
The finite-dimensionality of the space $\Vac_1\da([z],q,\pp;\vlm)$ can
be shown in a similar way as in [TUY].
The compatibility of
(D1)--(D4) implies
 that there exists a vector bundle $\tilde\Vac_1\da(\vlm)$
over a domain $U\subset T_N\times \C^*$ which has $\Vac_1\da([z],q,\pp;\vlm)$
 as a fiber at
$([z],q,\pp)\in U$, with the integrable connections defined by the
 differential equations (D2)--(D4). In particular the dimension of
the fiber $\Vac_1\da([z],q,\pp;\vlm)$ does not depend on $([z],q,\pp)$.

%
%
\medskip
{\bf 3.3 Restrictions of $N$-point functions to $V_\vlm$.}
\medskip
In this subsection we see that, as a consequence of the twisting,
an $N$-point function in genus 1
is determined from its restriction to $V_\vlm$ (Proposition 3.3.2).
We also give the characterization of $N$-point functions as $V_\vlm\da$-valued
 functions (Theorem 3.3.4).
\proclaim
{Lemma 3.3.1}
Let $\cals$ be the subspace of
 $\H_\vlm$ spanned by the vectors
$$
\rho_1(H_{-1})^k|v\> \ \ \ (|v\>\in V_{\vlm},\  k\in\Z_{\geq 0}).
$$
Then  for $(z,q,\pp)\in T_N\times\C^*$ such that $\pp\ne q^n\
(n\in\Z)$,
the natural map
$$
\cals\  \to \Vac_1([z],q,\pp;\vlm)
$$
is  surjective. In other words
the restriction map
$$
\Vac_1^\dagger([z],q,\pp;\vlm)\ \to\ \Hom_\C(\cals,\C)
$$
is injective.
\endproclaim

{\it Proof. }
For $\pp\neq q^n\ (n\in\Z)$, the space $\gh([z],q,\pp)$ is spanned by
 the following
$\g$-valued functions
$$\gather
H\otimes 1 ,\ H\otimes \left(\zeta(t/z_i,q)-\zeta(t/z_j,q)\right),\
H\otimes (\pat)^n\tP(t/z_j,q), \\
E\otimes (\pat)^n\sigma_{+}(t/z_j,q,\pp),\
F\otimes (\pat)^n\sigma_{-}(t/z_j,q,\pp)\ (i,j=1,\ldots,N,\ n=0,1,\ldots).
\endgather
$$
Note that  for each $i=1,\ldots,N$ and
$n\in\Z_{\geq 0}$, the function $(\pat)^n\tP(t/z_i)$ has
a pole of order $n+2$ and $(\pat)^n\sigma_\pm(t/z_i)$ has
a pole of order $n+1$ at $z_i$.
Let $\widetilde\cals$ be the subspace of $\H_\vlm$ spanned by the vectors
$$\rho_1(H_{-1})^{n_1}\cdots\rho_N(H_{-1})^{n_N}|v\>\ \ \
(n_1,\ldots,n_N\in\Z_{\geq0},\; |v\>\in V_\vlm).$$
Then we  can show that the natural map
$\widetilde\cals\to\Vac_1([z],q,\pp)$ is surjective,
 by induction with respect to
the filtration  $\{\F_\bullet\}$ on $\H_\vlm$.
Namely, for any given $|u\>$ of $\F_p\H_\vlm$ with
$|u\>\notin\widetilde\cals$, we can find an element $|u'\>$
in $\gh([z],q,\pp)\H_\vlm$
 such that
$$|u\>-|u'\>\in\F_{p-1}\H_\vlm$$

Now, to prove Lemma 3.3.1, it is sufficient to note (3.2.2).
  $\square$
\medskip

Let $\<\Phi|$ be an $N$-point function in genus 1 and
$|u\>$ be a vector in $\H_\vlm$.
By Lemma 3.3.1 we can express
$\<\Phi(z,q,\pp)|u\>$ as a combination of
$$\<\Phi(z,q,\pp)|\rho_1(H_{-1})^n|v\>\ \ (n\in\Z_{\geq 0}, |v\>\in V_\vlm).$$
Combining with (D4) we have the procedure to rewrite $\PH(z,q,\pp)|u\>$
as a combination of
$$
\left(\pap\right)^n\<\Phi(z,q,\pp)|v\>\ \ (n\in\Z_{\geq 0},|v\>\in V_\vlm)
\, .
$$
Furthermore it is easily seen from the proof of Lemma 3.3.1
that we need finitely many data for each $|u\>$:
\proclaim
{Proposition 3.3.2}
For $|u\>\in \F_p\H_\vlm$, there exist functions
$$a_{i,n}(z,q,\pp)\
(i=1,\ldots,\dim V_\vlm,\; n=1,\ldots,p)$$
 on $T_N\times \C^*$
such that
$$\PH(z,q,\pp)|u\>
=\sum_{i,n}a_{i,n}(z,q,\pp)\left(\pap\right)^n\PH(z,q,\pp)|b_i\>
$$
for any $\PH|\in\f_1(\vlm)$, where
$\{\; |b_i\>\; ;\; i=1,\ldots,\dim V_\vlm\; \}$ is a basis of $V_\vlm$.
\endproclaim
By $\f_1^r(\vlm)$ we denote the image of $\f_1(\vlm)$ in
$V_\vlm\da[T_N\times\C^*]$
under the restriction map to $V_\vlm$, which is injective by the above
proposition.

Next, as in the case of genus 0, we consider
the characterization of $\f_1^r(\vlm)$
in $V_\vlm\da[T_N\times\C^*]$.
First, we have the following.
\proclaim
{Proposition 3.3.3}
The restriction $\<\phi|$ of an $N$-point function satisfies
the following equations.
\medskip

(E1)
$$\sum_{j=1}^N \<\phi(z,q,\pp)|\rho_j(H)=0.$$

(E2) For each $j=1,\ldots,N$,
$$
\align
&(\l+2)\left(z_j\pazj+\varDelta_{\lm_j}\right)
\left(\Th(\pp,q)\<\phi(z,q,\pp)|\right)=\\
&\pap\left(\Th(\pp,q)\<\phi(z,q,\pp)|\right)\rho_j(H)
+\sum_{i\neq j}
\Th(\pp,q)\<\phi(z,q,\pp)|\Omega_{i,j}(z_j/z_i,q,\pp),
\endalign
$$
where
$$\align
&\Omega_{i,j}(t,q,\pp)=\\
\frac{1}{2}\zeta(t,q)&\rho_i(H)\rho_j(H)
+\sigma_+(t,q,\pp)\rho_i(F) \rho_j(E)
+\sigma_-(t,q,\pp)\rho_i(E) \rho_j(F).
\endalign
$$

(E3)
$$\align
&(\l+2)\paq \left(\Th(\pp,q)\<\phi(z,q,\pp)|\right)=\\
&\left(\pap\right)^2\left(\Th(\pp,q) \<\phi(z,q,\pp)|\right)
+\sum_{i,j=1}^N\Th(\pp,q)\<\phi(z,q,\pp)|
\Lambda_{i,j}(z_i/z_j,q,\pp).
\endalign
$$
Here
$$\align
&\Lambda_{i,j}(t,q,\pp)=
\frac{1}{4}\left(\zeta(t,q)^2-\wp(t,q)\right)\rho_i(H)\rho_j(H)\\
&+\omega_+(t,q,\pp)\rho_i(E)\rho_j(F)
+\omega_-(t,q,\pp)\rho_i(F)\rho_j(E),
\endalign
$$
where $\omega_\pm(t,q,\pp)$ denote the functions defined by
$$\omega_{\pm}(t,q,\pp)=\frac{1}{2}\left\{
\pa_t\sigma_\pm (t,q,\pp)+(\zeta(t,q)\pm\zeta(\pp,q))
\sigma_\pm(t,q,\pp)\right\},
$$
which are holomorphic at $t=1$.
\endproclaim
For the proof of Proposition 3.3.3, we refer the reader to [FW].

\remark
{Remark}
 The equation (E2) is derived by Bernard as a equation for the trace of
the vertex operators (see \S\S 3.4),
he also derived (E3) in a special case.
The equations (E2)(E3) are called the
 Knizhnik-Zamolodchikov-Bernard (KZB) equations in [Fe][FW].
\endremark
\medskip
Note that the system of equations (E1)--(E3)
is not holonomic since we have $j+2$ parameters
$\zN,q,\pp$,  but have only $j+1$ differential equations, which are compatible
each other.

The differential equations (E2) and (E3) are of order 1 with respect to
$z_j\ (j=1,\ldots,N)$ and $q$ respectively.
Hence to characterize $\f_1^r(\vlm)$ in $V_\vlm[T_N\times \C^*]$,
it is sufficient to obtain
equations which determine the $\pp$-dependence of the restricted
$N$-point functions
and they are obtained as follows.

Let $\PH|$ be an $N$-point function and $\<\phi|$
its restriction to $V_\vlm$.
We put $\M_\vlm\da=\Hom_\C(\M_{\lm_1}\otimes\cdots\otimes\M_{\lm_N}\,\C)$ and
regard $\PH|$ as an $\M_\vlm\da$-valued function.
Then as a special case of integrability condition,
 we have for each non negative integer $k$
$$\<\Phi|v_1\otimes\cdots\otimes F^k E_{-1}^{\l-2\lm_j+1}
{v}(\lm_j)\otimes \cdots\otimes v_N\>=0\tag 3.3.1$$
for any $|v_i\>\in V_{\lm_i}\ (i\neq j)$, where $|v(\lm_j)\>$ denotes
the highest weight vector in $\M_{\lm_j}$.

On the other hand, by Proposition 3.3.2
we can rewrite the left hand side of (3.3.1)
as a combination of
$$(\pap)^n\<\Phi|v\>=(\pap)^n\<\phi|v\>\ \ \
(n=0,1,\ldots,\l-2\lm_j+1, |v\>\in V_\vlm).$$
Now the equality (3.3.1) implies the differential equation
for $\<\phi|$ with respect to $\pp$ of order at most $\l-2\lm+1$.
We denote this differential equation by
$$\<\phi|v_1\otimes\cdots\otimes F^k E_{-1}^{\l-2\lm_j+1}
v(\lm_j)\otimes\cdots\otimes v_N\>=0.
$$

\proclaim
{Theorem 3.3.4}
The space  $\f_1^r(\vlm)$ coincides with the solution space
of the system of equations (E1)--(E4), where (E4) is given by
\medskip
(E4) For each $j=1,\ldots,N$ and nonnegative integer
$k\le\sum_{i=1}^N \lm_i+\l-2\lm_j+1$,
$$
\<\phi(z,q,\pp)|v_1\otimes\cdots\otimes
F^{k}E_{-1}^{\l-2\lm_j+1}
v(\lm_j)\otimes \cdots\otimes v_N\>=0\, ,
$$
for any $|v_i\>\in V_{\lm_i}(i\neq j)$.
\endproclaim
\demo
{Proof}
It is enough to prove
that the dimension of the solution space of the system (E1)--(E4) is not
larger than $\dim_\C\f_1(\vlm)=\dim_\C\f_1^r(\vlm)$.

Fix $(z,q)\in T_N$ and let $\<\phi(\pp)|=\<\phi(z,q,\pp)|$
be a $V_\vlm\da$-valued function on $\C^*$ which satisfies (E1) and (E4).
{}From $\<\phi(\pp)|$, we construct an element $\PH(\pp)|$ of
$\H_\vlm\da[\C^*]$
which satisfies
$$
\align
&\PH(\pp)|v\>=\<\phi(\pp)|v\>\ \hbox { for }|v\>\in V_\vlm,\tag i\\
&\PH(\pp)|\in\Vac_1\da([z],q,\pp;\vlm) \ \hbox{ for each }\pp\in\C^*,
\tag ii\\
&\pap\PH(\pp)|=\PH(\pp)|\frac{1}{2}H[\zeta(t/z_1) ],
\tag iii
\endalign
$$
through the steps below.
This means that we have the injective homomorphism from the solution space of
 (E1)--(E4) to the space of functions on $\C^*$ satisfying
 (ii) and (iii), and such space has the same dimension as
$\Vac_1\da([z],q,\pp;\vlm)$.
\medskip
{\it Step 1. }
We extend $\<\phi(\pp)|$ to a  $\M_\vlm\da$-valued function $\<\Phi(\pp)|$
on $\C^*$ in such  a way that $\<\Phi(\pp)|=\<\Phi(z,q,\pp)|$ satisfies
$$\align
&\<\Phi(\pp)|X[f]=0\ \ \ \hbox{for }X[f]\in\gh([z],q,\pp),\\
&\<\Phi(\pp)|\frac{1}{2}H[\zeta(t/z_1,q)]=\pap\<\Phi(\pp)|\,.
\endalign
$$
The well-definedness is proved by induction with respect to
 the filtration $\{\F_\bullet\ \}$  using Lemma 3.2.1 and
the compatibility condition (3.2.1).
\medskip
{\it Step 2. }
We show that $\<\Phi(\pp)|$ belongs to $\H_\vlm\da$, that is,
$$\<\Phi(\pp)|u_1\otimes\cdots\otimes
a\cdot E_{-1}^{\l-2\lm+1} v(\lm_j)\otimes \cdots\otimes u_N\>=0$$
for any $j=1,\ldots,N$, $|u_i\>\in\M_{\lm_i}$ and $a\in U(\widehat{\frak
p}_-)$.
This is reduced to (E4) also by induction.
\ $\square$
\enddemo
\medskip
In the case of $N=1$ we can write down the differential equations
(E4) explicitly as we will
see in \S4.
%
\medskip
{\bf 3.4 Sewing procedure.}
\medskip
In this subsection
we show that the $N$-point functions in genus 1 are given by the traces of
vertex operators and hence Bernard's approach is equivalent to ours.
For this purpose we construct an $N$-point function in genus 1
from an $N+2$-point function in genus 0.
This construction is known as the sewing procedure.

Fix $\mu\in P_\l$
and  $\vlm=(\lm_1,\ldots,\lm_N)\in (P_\l)^N$,
and consider a sequence of vertex operators
$\varphi_j(z_j)
: V_{\lm_j}\otimes\H_{\mu_{j}}\to \hat\H_{\mu_{j-1}}$
for some $\mu_{j-1},\mu_{j}\in P_\l$ with $\mu_0=\mu_{N}=\mu$.
For $|u\>=|u_1\otimes\cdots \otimes u_N\>\in\H_\vlm$, we put
$$\Phi_0(|u\>;z)=\hat\varphi_1(|u_1\>;z_1)
\hat\varphi_2(|u_2\>;z_2)\cdots\hat\varphi_N
(|u_N\>;z_N) : \H_\mu\to\hat\H_\mu,
$$
where $\hat\varphi_j(z_j)$ means the extended vertex operator
 in the sense of Proposition 2.3.3.
We define a $\H_\vlm\da$-valued function on $T_N\times\C^*$ by
$$\<\Phi_1(z,q,\pp)|u\>=\tr \left(\Phi_0(|u\>;z)\q\p\right)
\tag 3.4.1$$
for $|u\>\in\H_\vlm$.
\proclaim
{Proposition 3.4.1}
The element $\<\Phi_1|$ of $\H_\vlm[T_N\times\C^*]$
defined by (3.4.1) is an $N$-point function in genus 1.
\endproclaim
\demo
{Proof}
First we prove that
$\<\Phi_1|$ satisfies the condition (D1).%

Fix any $X\otimes f\in \gh([z],q,\pp;\vlm)$ and $|u\>\in\H_\vlm$,
 and put
$$
\<\Phi_1|X(t)|u\>dt= \tr\Phi_0(|u\>;z)X(t)\q\p dt.
$$
This is a holomorphic 1-form on $\C^*\setminus
\{\; q^nz_j\in\C^*\; ;\; n\in\Z,j=1,\ldots,N\; \}$.
Then by (2.3.1), what we should show is the following.
$$
\sum_{j=1}^N\mathop{\Res}_{t=z_j}f(t)\<\Phi_1|X(t)|u\>dt=0.\tag 3.4.2
$$
We have
$$
\align
f(t)\<\Phi_1|X(t)|u\>dt &= f(t)\tr X(t)\Phi_0(|u\>;z)\q\p dt \\
        &= f(t)\tr \Phi_0(|u\>;z) \q\p X(t)dt \\
        &= f(qt)\tr\Phi_0(|u\>;z) X(qt)\q\p d(qt)\\
        &= f(qt)\<\Phi_1|X(qt)|u\>d(qt),
\endalign
$$
 where we used the commutativity of vertex operators and currents, and
$$
f(t) \p (X(t)) \pp^{-\frac{H}{2}}=f(qt)X(t),\
q^{L_0} (X(t)) q^{-L_0} = X(qt)q.
$$
Therefore we have $f(t)\<\Phi_1|X(t)|u\>dt
\in H^0(\E_q,\omega_{\E_q}(\sum_{j=1}^N*[z_j]))$,
where $\omega_{\E_q}$ denotes the sheaf of 1-forms on $\E_q$.
This implies (3.4.2).
\medskip
Next we prove that
$\PH|$ satisfies the equation (D2)--(D4).
It is obvious that $\PH|$ satisfies (D2) from (2.3.2).
We give a proof of (D4). The equation (D3) is proved in a similar way.
We chose $(z,q)$ from
 the region $1>|z_1|>|z_2|>\cdots >|z_N|>|q|$, where
$\<\Phi_1|$ is a convergent power series.
Let $Z_r=\{|w|=r\}$ be a cycle with anticlockwise orientation.
We have
$$
\align
&2 \pi \sqrt{-1} \<\Phi_1|H[\zeta(t/z_1)]|u\>\\
&= \tr\int_{Z_{1}} \zeta(t/z_1)H(t)\Phi_0(|u\>;z)\q \p dt\\
&- \tr\int_{Z_{q}} \zeta(t/z_1)\Phi_0(|u\>;z)H(t)\q \p dt\\
&= \tr
\left\{
 \int_{ Z_{q} } \zeta(q^{-1}t/z_1)
-\int_{ Z_{q}    } \zeta(t/z_1)
\right\}
\Phi_0(|u\>;z)H(t)\q \p dt.
\endalign
$$
By $\zeta(t)=\zeta(q^{-1}t)-1$,
we conclude
$$\align
&\<\Phi_1|H[\zeta(t/z_1)]|u\>\\
&=\frac{1}{2\pi \sqrt{-1}}\tr \int_{Z_{q}}
\Phi_0(|u\>;z)H(t)\q \p dt\\
&=\tr\left(\Phi_0(|u\>;z)H \q \p \right).
\endalign
$$
This proves (D4). $\square$
\enddemo
By Proposition 3.4.1 we have the mapping from $\f_0(\mu,\vlm,\mu)$
to $\f_1(\vlm)$. We denote this mapping by $s_\mu$.
The following proposition follows from ``the factorization property''
proved in [TUY].
\proclaim
{Proposition 3.4.2}
The following map is bijective.
$$\mathop{\oplus}_{\mu\in P_\l}s_\mu:
\bigoplus_{\mu\in P_\l}\f_0(\mu,\vlm,\mu)\ \to\
\f_1(\vlm).\ \ \ \square$$
\endproclaim

By Proposition 2.3.2 and Proposition 3.4.2 we get the following.

\proclaim
{Theorem 3.4.3}
 The  space $\f_1^r(\vlm)$ is spanned by the functions
$$\hbox{Tr}_{\H_\mu}\Vo_1(z_1)\cdots\Vo_N(z_N)\q\p,$$
where $\Vo_j(z_j)\ (j=1,\ldots,N)$ is the vertex operator of type
$(\mu_{j-1},\lm_j,\mu_{j})$ for some $\mu_i\in P_\l\ (i=1,\ldots,N+1)$
with $\mu:=\mu_0=\mu_{N}$.
\endproclaim
\remark
{Remark}
The integral representations of the above functions are obtained in [BF].
\endremark

\subhead
\S4 Explicit formulas for 1-point functions in genus 1
\endsubhead
\medskip
In this section, we see how the system (E1)--(E4) determine
the 1-point function explicitly (Theorem 4.2.4).
 We also solve the system in a few cases in the last subsection.
\medskip
{\bf 4.1. The 1-point functions in genus 1.}
\medskip
Fix a weight $\lm$ and consider the set $\f_1^r(\lm)$ of restricted
1-point functions in
genus 1, which is, by Theorem 3.4.3,
spanned by the following $V_\lm\da$-valued functions:
$$\<\phi_\mu(z_1,q,\pp)|:=\tr \Vo(z_1)\q\p\ \ (\mu\in P_\l),$$
where $\Vo(z_1)$ is the vertex operator of type $(\mu,\lm,\mu).$
We
 put
$$L=\l-2\lm.$$
Note that
a nonzero vertex operator of type $(\mu,\lm,\mu)$ exists if and only if
$\lm$ and $\mu$ satisfy
$$\lm\in\Z,\ \frac{\lm}{2}\le \mu\le\frac{\lm+L}{2},$$
 and the vertex operators are unique up to constant multiples.
In particular we have
$$\dim_\C \f_1(\lm)=L+1.$$

As we have seen in Theorem 3.3.4, the restrictions of 1-point functions
 $\<\phi|$ are characterized by (E1)--(E4).
 The equation (E2) now implies
$$\<\phi(z_1,q,\pp)|=z_1^{-\varDelta_\lm}\<\phi(1,q,\pp)|.$$
Hence in the following we specialize $z_1=1$ and put
$$\<\phi(\pp,q)|=\<\phi(1,q,\pp)|.$$
The condition (E1) implies
$$\<\phi(\pp,q)|v\>=0$$
unless $|v\>\in\C|0_\lm\>$, where $|0_\lm\>$ is the weight 0 vector in
$V_\lm$ defined by
$$|0_\lm\>=\frac{1}{\lm!}F^\lm|v(\lm)\>.$$
We put
$$\phi_\mu(\pp,q)=\<\phi_\mu(\pp,q)|0_\lm\>=\Tr_\mu(|0\>_\lm;1)\q\p,$$
and identify $\f_1^r(\lm)$ with
 the space
spanned by the following function:
$$
 \phi_\mu (\pp,q)\ \ \  \mu=\plm\, .
$$
{}From the equation (E3) we immediately have the following heat equation.
\proclaim
{Proposition 4.1.1}
For $\phi \in \f_1^r(\lm)$,
$$
\align
&(\l+2)\paq (\Th(\pp,q)\phi(\pp,q))=\tag 4.1.1\\
&\left\{\left(\pap\right)^2
-\lm(\lm+1)\left( \wp(\pp,q)-2\frac{\paq\eta(q)}{\eta(q)}\right)
\right\}
(\Th(\pp,q)\phi(\pp,q)).
\endalign
$$
\endproclaim
\remark
{Remark}
The heat equations for 1-point functions are studied
by Etingof and Kirillov in more general cases: $\g=\frak{sl}(n,\C)$,
$V_\lm=S^{\lm}\C^n\ (\lm\in\Z)$, where $S^\lm$ denotes $\lm$-th
symmetric product and $\C^n$ the defining representation of $\g$ [EK].
\endremark

\medskip
{\bf 4.2 The differential equation with respect to $\pp$.}
\medskip

This subsection is devoted to write down
differential equations for
by $\phi\in \f_1^r(\lm)$ derived from (E4):
$$\<\phi|F^kE_{-1}^{L+1}|v(\lm)\>=0\ \ \ (0\le k\le \lm+L+1).$$
Among them  the only nontrivial equality is
the following:
$$\<\phi|F^{\lm+L+1}E_{-1}^{L+1}|v(\lm)\>=0,\tag 4.2.1$$
because other equalities fall into trivial by (E1).

To rewrite (4.2.1) as a differential equation with respect to $\pp$,
 we consider the following set of vectors in $\M_\lm$
$$
\left\{\ \uk = \frac{1}{(\lm+k)!}F^{\lm+k} E_{-1}^k|v(\lm)\>\ ;\  k=0,1,\ldots
\right\}_.
$$
Note that
$
|u^0\>=|0_\lm\>.
$
The following lemma plays a key role in the following discussions.
\proclaim
{Lemma 4.2.1}
For $k\in\Z_{\geq0}$, we have
$$\align
&\frac{1}{2}H[{\zeta}(z,q)]|u^k\>\ \equiv\ \tag$4.2.2$\\
-\frac{1}{2}|u^{k+1}\>+(k+\lm)&\zeta(\pp,q) |u^k\>
+k(L-k+1)\beta(\pp,q)
|u^{k-1}\>
\hbox{ mod }\gh([\vz],q,\pp)\M_\lm,
\endalign
$$
where $\beta(\pp,q)$ is given by
$$\beta(\pp,q)=\frac{\paq \Th(\pp,q)}{\Th(\pp,q)}
-3\frac{\paq \eta(q)}{\eta(q)}.
\tag$4.2.3$
$$
\endproclaim

{\it Proof. } First we have $$\frac{1}{2}H[{\zeta(z,q)}]|u^k\>\ =\
\frac{1}{2}( H_{-1}-2\al(q) H_1)|u^k\>
\tag$4.2.4$
$$ since $H_n|u^k\>=0$ for $n=0$ and $n\geq 2$.
The second term in the right hand side
 of $(4.2.4)$ can be calculated as follows:
$$
H_1|u^k\>=-2k(L-k+1)|u^{k-1}\>.\tag$4.2.5$
$$
This follows from
$$
\align
&H_1 F^{\lm+k}E_{-1}^k|v(\lm)\>=[H_1,F^{\lm+k}E_{-1}^k]|v(\lm)\>\\
&=-2(\lm+k)F^{\lm+k-1}F_1 E_{-1}^k|v(\lm)\>\\
&=-2k(L-k+1)(\lm+k)F^{\lm+k-1}E_{-1}^{k-1}|v(\lm)\>.
\endalign
$$

To rewrite the first term in the right hand side of $(4.2.4)$ as a linear
combination of $|u^n\>$'s, we use the equality
$$[E_{-1},F^{\lm+k+1}E_{-1}^k]=(\lm+k+1)H_{-1}F^{\lm+k}E_{-1}^k
+(\lm+k)(\lm+k+1)F_{-1}F^{\lm+k-1}E_{-1}^k.
$$
Thus we have
$$
\align
&(\lm+k+1)H_{-1}F^{\lm+k}E_{-1}^k|v(\lm)\>\\
&=E_{-1}F^{\lm+k+1}E_{-1}^k|v(\lm)\>-F^{\lm+k+1}E_{-1}^{k+1}|v(\lm)\>\\
&-(\lm+k)(\lm+k+1)F_{-1}F^{\lm+k-1}E_{-1}^k|v(\lm)\>\\
&\equiv
E\left[(z-1)^{-1}-\sigma_+(z)\right]F^{\lm+k+1}E_{-1}^k|v(\lm)\>
-F^{\lm+k+1}E_{-1}^{k+1}|v(\lm)\>\\ &\
-(\lm+k)(\lm+k+1)F\left[(z-1)^{-1}-\sigma_-(z)\right]F^{\lm+k-1}E_{-1}^k|v(\lm)\>
\ \hbox{ mod }\gh_\pp(\X_q)\M_\lm.
\endalign
$$
Since we can easily check
$$
\align
&E_nF^{\lm+k+1}E_{-1}^k|v(\lm)\>=0,\\
&F_nF^{\lm+k-1}E_{-1}^k|v(\lm)\>=0
\endalign
$$
for $n\geq2$,
we have
$$
\align
&E\left[(z-1)^{-1}-\sigma_+ (z)\right]F^{\lm+k+1}E_{-1}^k|v(\lm)\>=\\
&+\left(\zeta(\pp,q)-\frac{1}{2}\right)
EF^{\lm+k+1}E_{-1}^k|v(\lm)\>
+(\beta(\pp,q)-2\al(q))
E_1F^{\lm+k+1}E_{-1}^k|v(\lm)\>
\endalign
$$
and have a similar equality for
$F\left[(z-1)^{-1}-\sigma_{-}(z)\right]F^{\lm+k-1}E_{-1}^k |v(\lm)\>$.
Here we used (3.1.8).
By some more straightforward calculations, we have (4.2.2).
$\square$
\medskip
For $\<\Phi|\in\f_1(\lm)$, we put
$$
\vec{\Phi}=\,{}^t (\<\Phi|u^0\>,\ldots,\<\Phi|u^L\>).\tag 4.2.6
$$

Then by $\<\Phi|u^{L+1}\>=0$ and Lemma 4.2.1,
 we obtain the following differential equation for
$\vec{\Phi}$.
\proclaim
{Proposition 4.2.2}
For 
$\<\Phi|\in \f_1(\lm)$, we have
$$
\pap \vec{ \Phi}(\pp,q)\
=\ \J_{L+1}(\pp,q) \vec{\Phi}(\pp,q)\
+\ \lm\zeta(\pp,q) \vec{\Phi}(\pp,q) .\tag$4.2.7$
$$
Here, $\J_{L+1}$ is an $(L+1)\times (L+1)$ tri-diagonal matrix given by
$$
\J_{L+1} = \ \ \ \ \ \ \ \ \ \ \ \ \ \ \ \ \ \ \ \ \ \ \ \ \ \ \ \ \ \ \ \ \ \
\ \ \ \ \ \ \ \ \ \ \ \ \ \ \ \ \ \ \ \ \ \ \ \ \
\tag 4.2.8
$$
$$\pmatrix
\ \ \ 0 \  & -\frac{1}{2}\ \  \\
\\
1\cdot L\cdot \beta & \zeta\ \  &  \ -\frac{1}{2} \\
\\
\ & 2\cdot (L-1)\cdot \beta & \; \ 2\zeta & \ \ \ \ \ -\frac{1}{2} \\
\\
  &                          & \ \ldots & \ \ \ \ \ \ \ \ldots &\ldots \\
\\
\\
& & & (L-1)\cdot 2 \cdot \beta & \ (L-1)\zeta\  &
\ \ -\frac{1}{2}\ \ \\
\\
& & & & L\cdot 1\cdot \beta & L\zeta
\endpmatrix_,
$$
where the functions $\zeta(\pp,q)$ and $\beta(\pp,q)$ are given by
$(3.1.2)$ and $(4.2.3)$.
\endproclaim
It is remarkable that the equation (4.2.7) can be written in the
following form:
$$
\pap \left( \Th ^{-\lm} \vec{\Phi}\right)\
 =\ \J_{L+1} \left( \Th ^{-\lm} \vec{\Phi}\right).\tag$4.2.9$
$$
By Proposition 4.2.2, we can write $\<\Phi|u^k\>$ as a combination of
differentials of $\phi:=\<\Phi|u^0\>$ with respect to  $\pp$; e.g.
$$\align
\<\Phi|u^1\> &= -2\pap\phi\, , \\
\<\Phi|u^2\> &= -2\left(\pap-\zeta\right)\<\Phi|u^1\>
-2L\beta\phi\\
           &= 4 \left(\pap-\zeta\right)\pap\phi
-2L\beta\phi\, ,\\
&\hbox{etc}...
\endalign
$$
In general, we have the following lemma by $(4.2.9)$ and
simple calculations.
\proclaim
{Lemma 4.2.3}
For $k=1,\ldots,L+1$, we have
$$
\Th^{-\lm}\<\Phi|u^k\>\ =\ (-2)^k \DT\left[
\pap\cdot I_k-\J_{L+1}^{(k)}\right]
\left(\Th^{-\lm} \phi\right)\; ,\tag$4.2.10$
$$
where $I_k$ is the $k\times k$-identity matrix and
$\J_{L+1}^{(k)}$ is the $k\times k$-matrix given by
the first  $k\times k$  block  of $\J_{L+1}$:
$$
\J_{L+1}^{(k)}\ =
\pmatrix
0 & -\frac{1}{2} & \\
\\
L \beta & \zeta & -\frac{1}{2} \\
\\
& 2(L-1)\beta &\  2\zeta &-\frac{1}{2}\ \ \ \ \ \ \ \ \ \ \\
\\
\\
& &\ \ \cdots &\cdots&\cdots\ \ \ \ \ \ \\
\\
\\
 & & & (k-1)(L-k+2)\beta & (k-1)\zeta
\endpmatrix_.
$$
Here, for an $n\times n$-matrix $A =(a_{i,j})$ with elements in some,
possibly non-commutative, ring, $\DT A$ is defined inductively
as follows:
$$\align
&\DT A\ =\ a_{1,1} \ \hbox{ for }n=1,\\
&\DT A\ =\ \DT A_{1,1} \cdot a_{1,1}\ -\ \DT A_{1,2} \cdot a_{1,2}
\ +\ \cdots\ +(-1)^{n-1} \DT A_{1,n} \cdot a_{1,n}\, ,
\endalign
$$
where $A_{i,j}$ is the matrix given by removing the $i$-th row
and $j$-th column from $A$.
$\square$
\endproclaim

Through this lemma,
we can rewrite (4.2.1) explicitly
as a differential equation for $\phi\in\f_1^r(\lm)$ of order $L+1$
with respect to $\pp$. Combining with (4.1.1) we have the following.
\proclaim
{Theorem 4.2.4}
The space $\f_1^r(\lm)$ coincides with the solution space
of the following system of differential equations.
$$
\align
&(\l+2)\paq \left(\Th(\pp,q)^{-\lm}\phi(\pp,q)\right)=\tag$F1$\\
&\left\{\left(\pap\right)^2+2(\lm+1)\zeta(\pp,q)\pap
-L(\lm+1)\frac{\paq \Th(\pp,q)}{\Th(\pp,q)}
\right\}\left(\Th(\pp,q)^{-\lm}\phi(\pp,q)\right).\\
&\DT \left[\pap\cdot I_{L+1}
-\J_{L+1}(\pp,q)
\right]
\left(\Th(\pp,q)^{-\lm}\phi(\pp,q)\right)\
 =\ 0\, .\tag$F2$
\endalign
$$
\endproclaim
\medskip
\remark
{Remark}
(i) It is easy to see directly that
the solution space of (F1)(F2) is $(L+1)$-dimensional.

(ii) For $\lm=0$,  the vertex operator
$\varphi_\mu(|0\>_0;z)$ is equal to the identity operator
on $\H_\mu$ up to a constant multiple.
Thus
the 1-point function $\phi_\mu$ is nothing but
the character
$$
\chi_\mu^{(\l)}(\pp,q)=\hbox{Tr}_{\H_\mu}\q\p=
\frac{\Th_{2\mu+1,\l+2}(\pp,q)-\Th_{-2\mu-1,\l+2}(\pp,q)}
{\sqrt{-1}\Th(\pp,q)},
$$
where $\Th_{m,k}(\pp,q)$ is the theta function of level $k$ defined by
$$
\Th_{m,k}(\pp,q)=\sum_{n\in \Z+\frac{m}{2k}}q^{kn^2}\pp^{kn}.
$$
In the case of  $\l=1,2$, the system
(F1)(F2) coincides
 with the one obtained in [EO2].
\endremark
\medskip
We can easily solve (F2) by noting
the above remark (ii).
\proclaim
{Proposition 4.2.5}
For $\lm\in P_\l$ and $q\in D^*$, the functions
$$
\Th(\pp,q)^{\lm}\chi_{\mu}^{(\l-2\lm)}(\pp,q)\ \ \
\left(\mu=0,\frac{1}{2},\ldots,\frac{\l-2\lm}{2}\right)
$$
form a basis of the solution space of (F2). $\square$
\endproclaim

\medskip
{\bf 4.3. Some solutions.}
\medskip
%
In this subsection we determine the trace of vertex operators
explicitly when $L=\l-2\lm\le 1$, by solving the differential equations
 (F1) and (F2).
\medskip
{\it Case } $L=0:$

In this case the space $\f_1^r(\lm)$ is spanned by the single function
$$\phi_{\frac{\lm}{2}}(\pp,q)
=\hbox{Tr}_{\H_{\frac{\lm}{2}}}\varphi(|0\>_{\frac{\lm}{2}};1)\q\p.
$$
On the other hand, by Proposition 4.2.5, any solution of (F2) is given
in the following form:
$$
\phi(\pp,q)=a(q) \Th(\pp,q)^{\lm} \chi_0^{(0)}(\pp,q)=a(q) \Th(\pp,q)^{\lm}
$$
with some function $a(q)$, and the equation (F1) now implies
$\pa_q a(q)=0.$
Therefore, we have
$$
\phi_{\frac{\lm}{2}}(\pp,q)=\Th(\pp,q)^\lm\tag 4.3.1
$$
under the appropriate normalization.
\medskip
{\it  Case } $L=1:$

The space $\f_1^r(\lm)$ has dimension 2 and  it is spanned by
$$
\phi_{\mu}(\pp,q)=\Tr(|0\>_{\mu};1)\q\p\ \ \
\left(\mu=\frac{\lm}{2},\frac{\lm+1}{2}\right).
$$
On the other hand, by substituting
$$
a_0(q) \Th(\pp,q)^{\lm}\chi_0^{(1)}(\pp,q)+a_1(q) \Th(\pp,q)^{\lm}
\chi_{\frac{1}{2}}^{(1)}(\pp,q)
$$
for $\phi(\pp,q)$ in (F1), and using (F1) for $L=1,\lm=0$,
we find that the functions
$$
\eta(q)^{-\frac{\lm}{2\lm+3}} \Th(\pp,q)^{\lm} \chi_\nu^{(1)}(\pp,q)\ \ \
\left(\nu=0,\frac{1}{2}\right)
$$
are solutions of the system.
By comparing the exponents of $q$, we conclude
$$
\split
\phi_{\frac{\lm}{2}}(\pp,q)=\;
&\eta(q)^{-\frac{\lm}{2\lm+3}}\Th(\pp,q)^\lm \chi_0^{(1)}(\pp,q),\\
\phi_{\frac{\lm+1}{2}}(\pp,q)=
\;&\eta(q)^{-\frac{\lm}{2\lm+3}}\Th(\pp,q)^\lm \chi_{\frac{1}{2}}^{(1)}(\pp,q).
\endsplit\tag 4.3.2
$$

\par
\medskip
{\bf Acknowledgements}
\medskip
I would like to thank T. Miwa
for stimulating encouragement and inspiring discussions.
I am also grateful to M. Kashiwara, H. Ooguri, A. Tsuchiya and K. Ueno for
useful suggestions and advice.


\Refs
\widestnumber\key{W-W}


\ref
\key Be1 \by D. Bernard
\paper On the Wess-Zumino-Witten models on the torus
\jour Nucl. Phys. \vol B303 \yr 1988 \pages 77--93
\endref

\ref
\key Be2 \by D. Bernard
\paper On the Wess-Zumino-Witten models on Riemann surfaces
\jour Nucl. Phys. \vol B309 \yr 1988 \pages 145--174
\endref

\ref
\key BF \by D. Bernard and G. Felder
\paper Fock representations and BRST cohomology in SL(2) current algebra
\jour Commun. Math. Phys. \vol 127  \yr 1990 \pages 145--168
\endref

\ref
\key BPZ \by A.A. Belavin, A.M. Polyakov and A.B. Zamolodchikov
\paper Infinite dimensional conformal symmetry in two-dimensional quantum field
theory
\jour Nucl. Phys. \vol B241 \yr 1984 \pages 333--380
\endref


\ref
\key EK \by P. Etingof and A. Kirillov, Jr.
\paper On the affine analogue of Jack's and Macdonald's polynomials
\jour Yale preprint (1994), hep-th/9403168, to appear Duke Math. J.
\endref

\ref
\key EO1 \by T. Eguchi and H. Ooguri
\paper Conformal and current algebras on a general Riemann surface
\jour Nucl.Phys. \vol B282 \yr 1987  \pages 308--328
\endref


\ref
\key EO2 \by T. Eguchi and H. Ooguri
\paper Differential equations for characters of Virasoro and affine Lie
algebras
\jour Nucl. Phys.\vol B313 \yr 1989 \pages 482--508
\endref


\ref
\key Fe \by G. Felder
\paper Conformal field theory and integrable systems associated to elliptic
curves
\jour to appear in th Proceeding of the International Congress of
Mathematicians, Zurich \yr 1994
\endref


\ref
\key FW \by G. Felder and C. Wieczerkowski
\paper Conformal blocks on elliptic curves and the
Knizhnik-Zamolodchikov-Bernard equations
\jour hep-th/941104\yr 1994
\endref


\ref
\key Ka \by V. Kac
\book Infinite dimensional Lie algebras
\publ Cambridge University Press., Third edition \yr 1990
\endref


\ref
\key KZ \by V.Z. Knizhnik and A.B. Zamolodchikov
\paper Current algebra and Wess-Zumino models in two dimensions
\jour Nucl. Phys. \vol B247 \yr 1984 \pages 83--103
\endref


\ref \key TK \by A. Tsuchiya and Y. Kanie
\paper Vertex operators in conformal field theory on $\P^1$ and monodromy
representations of braid group
\jour Adv. Stud. in Pure Math.\vol 16 \yr 1988 \pages 297
\endref


\ref
\key TUY \by A. Tsuchiya, K. Ueno and Y. Yamada
\paper Conformal field theory on universal family of stable curves with gauge
symmetries
\jour Adv. Stud. in Pure Math. \vol 19 \yr 1989 \pages 459--566
\endref

\endRefs
\enddocument
\end